\documentclass[12pt]{article}
\usepackage{epsfig}
\usepackage{graphicx}

\usepackage{times}

\newenvironment{sciabstract}{%
\begin{quote} \bf}
{\end{quote}}

\newcounter{lastnote}

\title{Ultrastrong coupling of the cyclotron transition of a two-dimensional electron gas to a THz metamaterial}

\author
{G. Scalari$^{1,\ast}$, C. Maissen$^1$, D. Tur\v {c}inkov\'a$^1$, D. Hagenm\"uller$^2$, \\
S. De Liberato$^2$, C.Ciuti$^2$, C. Reichl$^3$, D. Schuh$^4$, \\
W. Wegscheider$^3$, M. Beck $^1$, J. Faist$^1$\\
\\
\normalsize{$^1$ Institute of Quantum Electronics, Eidgen\"ossische  Technische Hochschule Z\"urich, Switzerland}\\
\normalsize{$^2$Laboratoire Mat\'eriaux et Ph\'enom\`enes Quantiques, }\\
\normalsize{Universit\'e Paris Diderot-Paris 7 and CNRS, Paris, France}\\
\normalsize{$^3$Laboratory for Solid State Physics,  Eidgen\"ossische  Technische Hochschule Z\"urich, Switzerland}\\
\normalsize{$^4$ Institut f\"ur Experimentelle und Angewandte Physik, Universit\"at Regensburg, Germany}
\\
\normalsize{$^\ast$E-mail:  scalari@phys.ethz.ch}
}
\date{}

\begin{document} 


\baselineskip24pt


\maketitle


\begin{sciabstract}
Artificial cavity photon resonators with ultrastrong light-matter interactions are attracting  interest both in semiconductor and superconducting systems, due to the possibility of manipulating the cavity quantum electrodynamic ground state with  controllable physical properties. We report here experiments showing ultrastrong light-matter coupling in a terahertz  metamaterial where the cyclotron transition of a high mobility two-dimensional electron gas  is coupled to the photonic modes of an array of electronic split-ring resonators.
 We observe a normalized coupling ratio $\frac{\Omega}{\omega_c}=0.58$ between the vacuum Rabi frequency $\Omega$ and the cyclotron frequency $\omega_c$.   Our system appears to be scalable in frequency and could be brought to the microwave spectral range  with the  potential of strongly controlling  the  magnetotransport properties of a high-mobility 2DEG.
\end{sciabstract}

%
%
%
%
%
%

Enhancement and tunability of light-matter interaction is crucial for fundamental studies of cavity quantum electrodynamics (QED) and for applications in classical and quantum devices \cite{Haroche:RMP:01, Walraff:Nat:04, Christopoulos:prl:2007,Hennessy:Nat:07}  . The coupling between one cavity photon and one elementary electronic excitation is quantified by the vacuum Rabi frequency $\Omega$. The non-perturbative strong light-matter coupling regime is achieved when $\Omega$  is larger than the loss rates of the photons and electronic excitations. Recently,  growing interest has been generated by the ultrastrong coupling regime\cite{Ciuti:PRB:05:115303-1, Ciuti:pra:2006, DeliberatoPRL2007, Devoret2007, Bourassa:pra:2009, muraev:prb:2011,Schwartz:prl:2011,natafprl2010}, which is obtained when the vacuum Rabi frequency becomes an appreciable fraction of  the unperturbed frequency of the system $\omega$. In such a regime,  it is possible to modify the ground and excited state properties obtaining non-adiabatic cavity QED effects \cite{Ciuti:PRB:05:115303-1}.
Experimental progress has been achieved in two different solid-state systems : (i) microcavities embedding doped quantum wells\cite{dini:PRL:2003,anapparaPRB, Guenter:Nature:2009,todorov:PRL:2010,geiser:APL:2010}, where the active electronic transition is between quantized subbands in the well; (ii) superconducting quantum circuits in transmission line resonators\cite{Niemczyk-NATPHYS-2010,fornprl2010}, where the photon field is coupled to artificial two-level atoms obtained with Josephson junctions. 

We present experimental results on a new system, namely a high-mobility two-dimensional electron gas (2DEG)  coupled to terahertz (THz) metamaterial resonators. The photon mode is coupled to the magnetic cyclotron transition of the 2DEG, obtained by applying a magnetic field perpendicular to the plane of the quantum wells  (Fig.1(A)).
  The cyclotron frequency is expressed by $\omega_c= \frac{e B}{m^*}$ where $B$ is the  applied magnetic field, \textit{e} is the elementary charge and $m^*$ represents the electron effective mass.  This highly controllable system is ideal for the study  of strong coupling because the material excitation can be continuously tuned by changing the value of the applied magnetic field. The key physical aspect to highlight is the dependence of the optical dipole moment $d$ for a cyclotron transition on the cyclotron orbit length. The dipole $d$ scales as $d \sim e l_0 \sqrt{\nu}$ \cite{Hagenmuller:2010p1619}, where  $l_0=\sqrt{\hbar/eB}$ is the magnetic length and $\nu=\rho_{\rm 2DEG} 2\pi l^{2}_{0}$ is the filling factor of the 2DEG , being $\rho_{\rm 2DEG}$ the electron areal density. This proportionality of the dipole with respect to  $l_0$ allows to have gigantic dipole moments as soon as the cyclotron transition can be resolved. 
 According to theoretical calculations valid for integer filling factors  and for an optimized resonator geometry, the coupling ratio is expected to scale as $\frac{\Omega}{\omega_c} \sim \sqrt{\alpha n_{\rm QW} \nu}$  where $\alpha$ is the fine structure constant and $n_{\rm QW}$ is the number of 2DEGs  \cite{Hagenmuller:2010p1619}. 
For high filling factors, this coupling ratio is predicted to assume values even larger than unity  (corresponding to transitions in the microwave range).  We use high-mobility 2DEG based on GaAs material system \cite{Umansky:JCG:09:1658}
and we realize our experiments in the THz region of the electromagnetic spectrum.  These frequencies, for our material system,  correspond to magnetic fields of the order of a few Tesla and optical experiments are conducted using broadband THz pulses generated with ultrafast lasers \cite{SMITH:JQE:88:1255}. A THz-TDS system (bandwidth $0.1-3 \,{\rm THz}$) \cite{GRISCHKOWSKY:JOSAB:90} is coupled to a split-coil superconducting magnet to probe sample transmission \cite{supportingmaterial}.

Our THz metamaterial integrates the 2DEG with a metasurface of electronic split-ring resonators (see Fig.1(A)). These resonators  \cite{SchurigSci:06:977,Chen:Nat:2006,padilla:prl:2006} exhibit electric field enhancement over strongly subwavelength volumes \cite{walther:science:2010}, making them ideal candidates to reach extreme couplings  in the Mid-Ir and THz range where long wavelength radiation has to interact with quantum well systems  typically extending over length of some micrometers \cite{Shelton:nanolett:2011}. Moreover, the enhanced in-plane (x-y) electric field couples efficiently to the cyclotron transition when the magnetic field is applied perpendicularly to the plane of the layers and parallel to the wavevector of the incident THz pulse (see Fig.1(A)). Resonators  were deposited on top of the 2DEG by conventional photolitography, metallization with Ti/Au ($5/250 \,{\rm nm}$)  and lift-off technique.

At zero magnetic field we observe two resonances m$_1$, m$_2$ whose origin is qualitatively different: the lowest frequency mode ($f_1\approx 0.9\,{\rm THz}$ ) is attributed to the  LC resonance, where counterpropagating currents circulate in the inductive part and the electric field is enhanced mainly in the capacitor gap \cite{Chen:Nat:2006}. The second mode ($f_2 \approx2.3 \,{\rm THz}$) is attributed to the "cut wire" behavior where a $\lambda/2$ kind of resonance is excited along the sides of the metaparticle \cite{padilla:prl:2006}.   These values correspond well to simulations with 3D FE modeling (see S1 in \cite{supportingmaterial}).
The two modes of the split-ring resonator also have different transverse wavevectors, as it is  evident looking at the different field distributions.
 The presence of conductive layers underneath alters the frequency and the quality factor of these resonances.

We observe a value of $Q_{1\rm THz}^{\rm (ins)} \simeq 4.3$ for an insulating substrate and  $Q_{1 \rm THz}^{\rm (2DEG)} \simeq 3.1$ when the resonator is deposited on top of the single 2DEG sample. The values for the second resonance results less affected  yielding  $Q_{2.3 \rm THz}^{\rm (ins)} \simeq 5.3$ and $Q_{2.3 \rm THz}^{\rm (2DEG)} \simeq 5.3$ (a more detailed analysis can be found in S1 of \cite{supportingmaterial} ). 
It is important to highlight  that,  in contrast to atomic systems, we can realize strong light-matter coupling physics with resonators displaying extremely low quality factors. The giant value of the  coupling constant $\Omega \sim \sqrt{n_{\rm QW} \rho_{\rm 2DEG}}$ typical of intersubband systems \cite{Ciuti:PRB:05:115303-1} together with the high electric field enhancement of sub-wavelength metallic resonators (V$_{cav}\simeq 8 \times 10^{-17}$ m$^3$ in our case ) allow the observation of cavity polaritons in a system where both components are in principle highly dissipative.

 In the data reported in Fig.2(A) we observe the evolution of the sample transmission $\vert t \vert$ $=\vert \frac{E_{\rm Meta} (B)}{E_{\rm 2DEG}(0)}\vert$  as a function of the applied magnetic field (normalized to the electric field $E_{\rm 2DEG}(0)$ of the reference 2DEG wafer at $B=0 \,{\rm T}$) . One 2DEG ($n_{\rm QW}=1$, doping density $\rho^{(1)}_{\rm 2DEG}=3.2 \times 10^{11} \,{\rm cm^{-2}}$) is used as an active medium and  placed $100  {\rm \,nm}$ below the surface: its cyclotron resonance can couple to the resonator modes. 
  As the magnetic field is swept a profound modification of the sample transmission is observed. The possibility to  tune in a continuous way the material excitation allows to follow the evolution of polaritonic states as the system is driven from the uncoupled regime to the strongly coupled one  \cite{tignon:prl:1995}. We  observe two successive anticrossings when the cyclotron energy matches the first and the second resonator modes. 

 In Fig. 2(B) we extract the positions of the minima of sample transmission and  plot the dispersion curves for the polariton eigenvalues as a function of the magnetic field. 
  The curves are calculated using a full quantum mechanical treatment of
the system, obtained generalizing the theory described in Ref.\cite{Hagenmuller:2010p1619} to
the case of a zero-dimensional resonator exhibiting two modes with
different transverse wavevectors \cite{supportingmaterial}.  Following a
bosonization procedure, we have derived the different contributions to
the total Hamiltonian and have diagonalized it using the Hopfield-Bogoliubov method. It is worth mentioning that the ideal resonator we have considered in the analytical treatment is different from the real split-ring one. However, we emphasize that this would introduce only a form factor in the matrix element calculation.

 In order to fit the experimental data, we need to know the resonator modes frequencies ($f'_1$ and $f'_2$ ) as well as the strength of their couplings ($\Omega_1$ and $\Omega_2$ ). For each cavity mode we assumed the asymptotic value of the corresponding  lower polariton branch to coincide with the frequency of the unloaded resonator ($f'_1=0.83$ THz and $f'_2=2.26$ THz)  \cite{Hagenmuller:2010p1619,supportingmaterial} . The coupling strength  $\frac{\Omega}{\omega}$ for the two modes can not be directly measured, we thus applied a best fit procedure following the least square method, analogously to what done in Ref. \cite{anapparaPRB} (more details in S3 of \cite{supportingmaterial}).
The minimal error \cite{supportingmaterial} is obtained for $\frac{\Omega_1}{\omega_1}=0.17$ and $\frac{\Omega_2}{\omega_2}=0.075$ (where $\omega_1=2\pi f'_1$ and $\omega_2=2\pi f'_2$).  As expected the coupling strength scales monotonously with $\nu$ :  for the measured density $\rho^{(1)}_{\rm 2DEG}$, we have  $\nu(B)=\nu(2\,T) \simeq 6.5$ for the first mode and $\nu(B)=\nu(5.5T)\simeq 2.4$ for the second one.

 
 To increase the coupling strength,  we kept the resonator geometry and hence the frequency constant, and  increased the effective number of carriers in the system. A new sample  was prepared with $n_{\rm QW}=4$ wells  and a doping density per well  $\rho^{(4)}_{\rm 2DEG}=4.45 \times 10^{11} \,{\rm cm^{-2}}$ (see scheme in Fig. 1(B) and  materials and methods in  \cite{supportingmaterial}). Sample transmission as a function of the applied magnetic field  (Fig.3(A))  shows that the system is driven deeply into the ultrastrong coupling regime. The polaritonic line widths display narrowing as the low quality cavity mode is mixed with the cyclotron resonance (Fig.3(B)).
Following the fitting procedure previously described we observe a value of $\frac{\Omega_1}{\omega_1}=0.36$ for the first resonance. Indeed, the effects of the anti-resonant terms of the light-matter Hamiltonian start becoming relevant when the dimensionless ratio $\frac{\Omega}{\omega}$ is of the order of $0.1$\cite{anapparaPRB}.  Due to the increased doping, the filling factor in the region of the anti crossing is $\nu \simeq 9$. As expected, the coupling ratio scales with $\sqrt{\rho_{\rm 2DEG}n_{\rm QW}}$ and for the two samples  at the same resonant frequency we obtain experimentally $\frac{\left(\frac{\Omega_1}{\omega_1}\right)_{n_{\rm QW}=4}}{\left(\frac{\Omega_1}{\omega_1}\right)_{n_{\rm QW}=1}}=\frac{0.36}{0.17}=2.11$ and theoretically $\sqrt{\frac{4 \rho^{(4)}_{\rm 2DEG}}{\rho^{(1)}_{\rm 2DEG}}}=2.35$. The small discrepancy can be attributed to the different coupling of the quantum wells which do not experience the same electric field of the resonator:  the generalized expression of the coupling ratio is calculated in the case when all the wells are coupled in the same way to the resonator's field \cite{Hagenmuller:2010p1619}.

By further scaling the resonator frequency down to  $f\simeq 500$ GHz employing a slightly modified geometry (see inset Fig.3(D), section S2 and Fig. S5 of  \cite{supportingmaterial} ) and employing the sample with n$_{QW}=4$ quantum wells we could probe the regime where the polariton splitting at the anticrossing $2\hbar \Omega$ is larger than the bare cavity photon energy. In Fig.3(D) we report the positions of the minima of the sample transmission for the case of f=500 GHz resonator together with the fitted dispersion curves:  we measure a normalized ratio $\frac{\Omega_1}{\omega_1}=0.58$ for a filling factor of $\nu(1.2 T)\simeq15.2$ which corresponds to $2\hbar \Omega \simeq 1.2 \omega_c$ (see also Fig.S6 in \cite{supportingmaterial} ).

 The generalization of the theory developed in Ref. \cite{Hagenmuller:2010p1619} accounts for the depolarization shift in presence of a magnetic field (magnetoplasmon) originating from the long-wavelength part of the Coulomb interaction. We found that the renormalization of the cyclotron transition frequency is too small to allow the experimental resolution of the magnetoplasmon branch (see Figs 2(A) and 3(A)) in our experimental parameter regime including the small wavevector condition $q l_{0} \ll 1$ which is always satisfied since dealing with optical wavevectors.

We have observed ultrastrong light matter coupling in a composite THz metamaterial measuring a normalized coupling ratio $\frac{\Omega}{\omega_c}=0.58$. The impact of our results has to be considered also in the  perspective of a  change in the  DC transport properties of the 2DEG, in analogy with what already observed  by direct irradiation at lower frequencies  \cite{Mani:Nat:02:646}.

These results should lead  to the scaling of the frequency to lower values and to an increase of effective density to further enhance the coupling strength.

\newpage

\begin{figure}[h]
\begin{center}
\includegraphics[width=100mm]{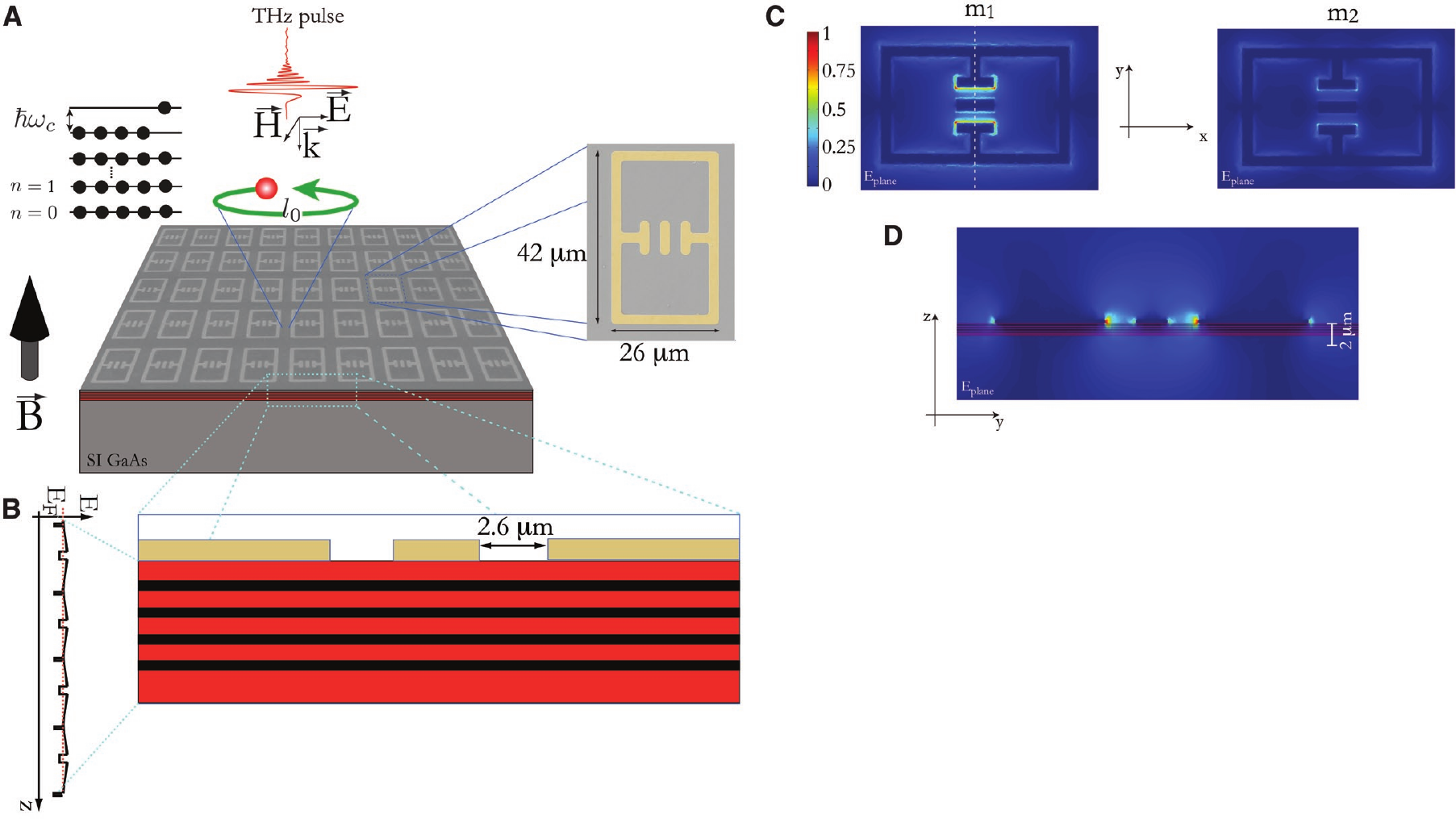}
\caption{(A):  Schematic of the composite metamaterial used in our experiment together with the experimental arrangement showing the polarizing static magnetic field, the wavevector and the polarization of the incident broadband THz pulse, the Landau level arrangement and the semiclassical representation of a cyclotron orbit of magnetic length $l_0$. A metasurface composed by LC-metaparticles  with a design  similar to Ref.\cite{Ohara2007} is deposited on top of the semiconductor. SEM picture of one metaparticle: the split gap of the capacitance elements is  $2.6 \,{\rm \mu m}$.   (B):  The bandstructure of the multi 2DEG system is schematized together with the quantum well position (not to scale) and (C) x-y spatial distribution of the in-plane electric field (E$_{plane}$=$\sqrt{\vert E_x\vert^2+\vert E_y \vert^2}$) calculated with a finite element commercial software is shown for the two observed modes m$_1$ and m$_2$. (z=100 nm below the semiconductor surface surface). (d): Intensity for E$_{plane}$ in the yz plane  for the low frequency mode m$_1$ (cut along the white dashed line in  (C) m$_1$ ).} 
\end{center}
\end{figure}

\begin{figure}[h]
\begin{center}
\includegraphics[width=130mm]{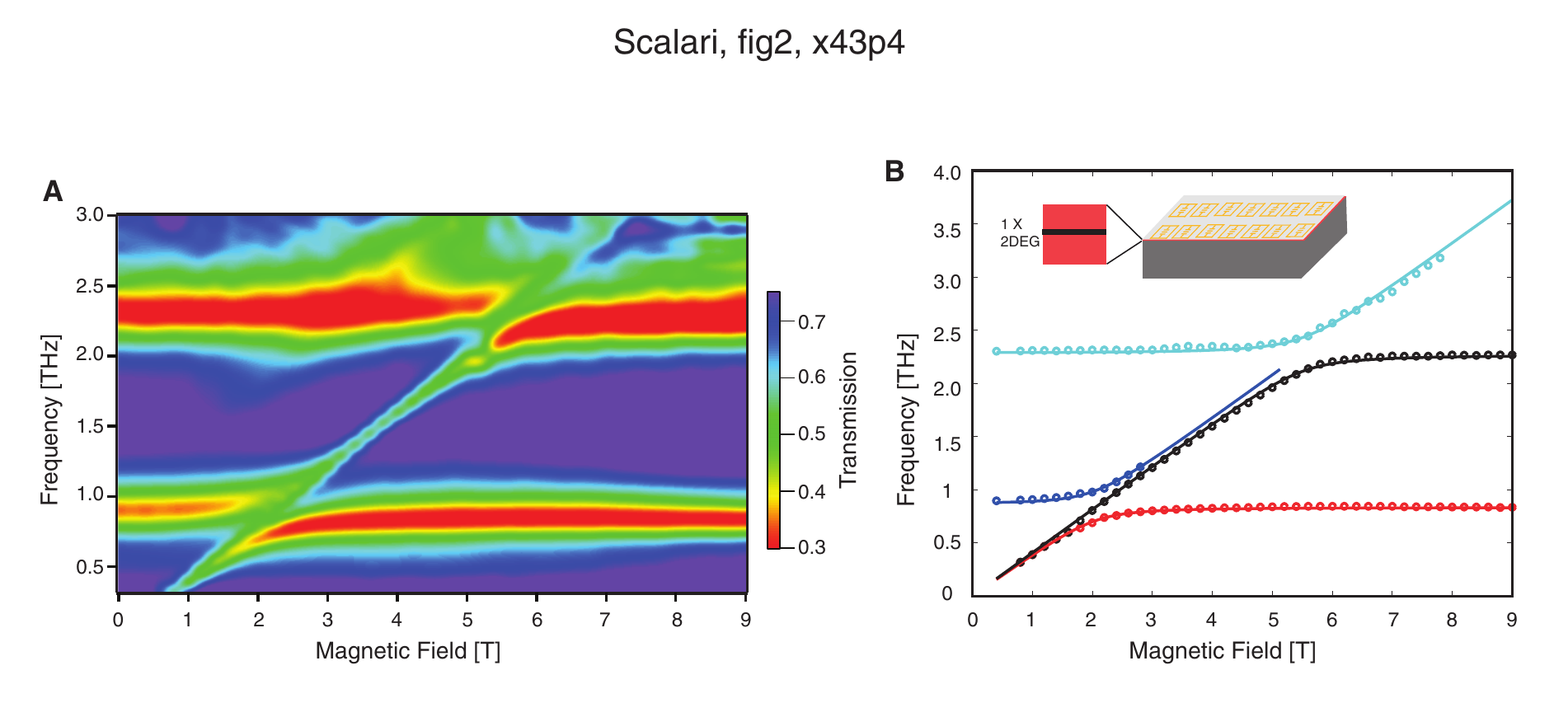}
\caption{((A): Transmission $\vert t \vert$ of the sample ($n_{\rm QW}=1$) as a function of the applied magnetic field $B$. The reference is a plain 2DEG sample without resonators on top and the measurement is performed at T=$2.2\,{\rm K}$. (B): Best fit with the extracted transmitted minima positions for the two different transverse modes of the electronic split ring  resonator; the fitting parameter is $\frac{\Omega}{\omega}$.} 
\end{center}
\end{figure}

\begin{figure}[h]
\begin{center}
\includegraphics[width=150mm]{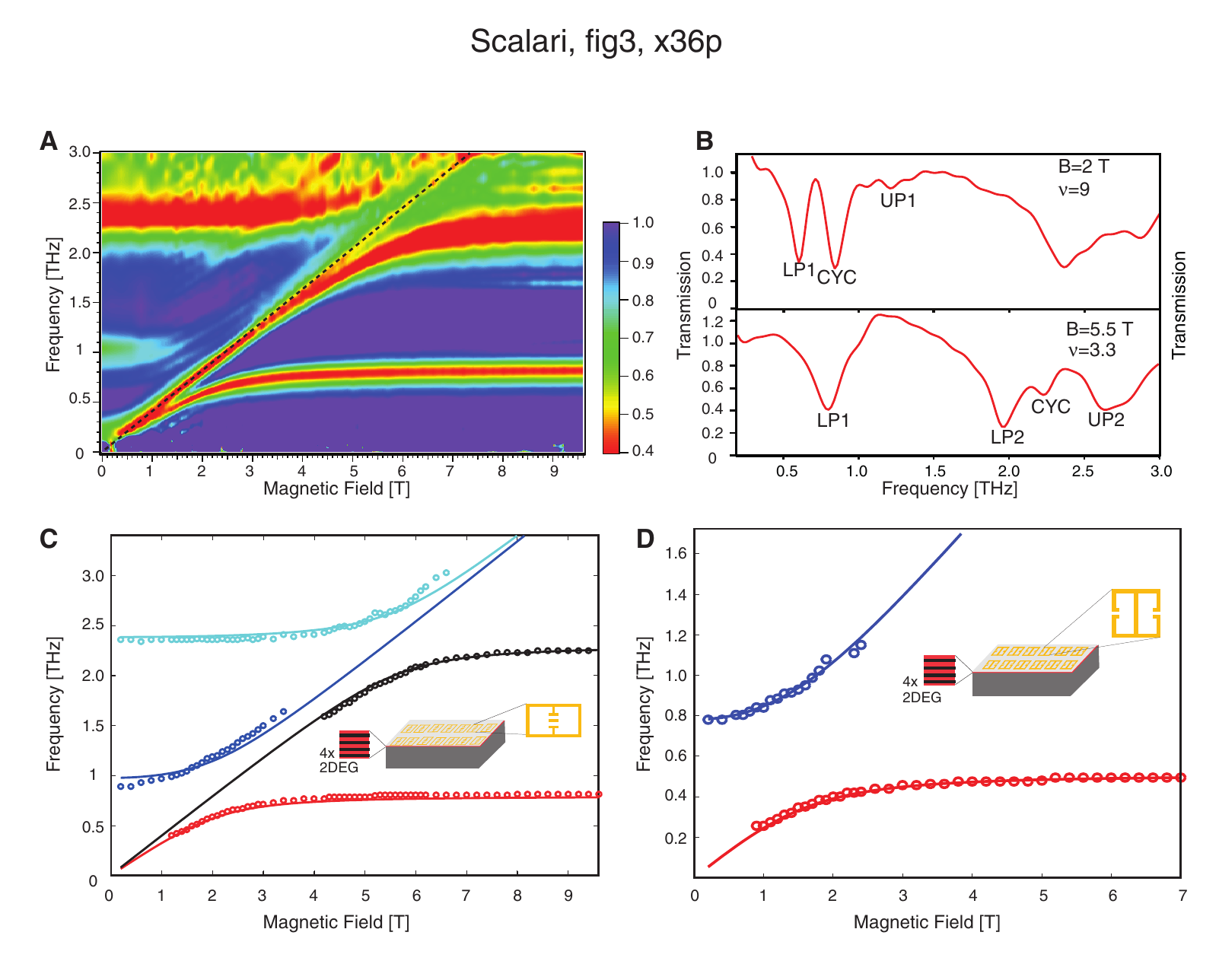}
\caption{(A): Transmission $\vert t \vert$ of the sample ($n_{\rm QW}=4$)  as a function of the applied magnetic field. The reference is a plain 2DEG sample without resonators on top and the measurement is performed at T=$10 \,{\rm K}$. The black dotted line highlights the cyclotron signal coming from the uncoupled material which is left in-between the resonators. (B): Sections in the two anticrossing regions  for the sample transmission. (C):  Best fit with the extracted transmitted minima positions for the two orthogonal modes of the split ring electronic resonator; the fitting parameter is $\frac{\Omega}{\omega}$. (D): Best fit with the extracted transmitted minima positions for the f=500 GHz resonator and $n_{\rm QW}=4$ measured at T=$10 \,{\rm K}$ ; the fitting parameter is $\frac{\Omega}{\omega}$. Inset: scheme of the 500 GHz resonator. } 
\end{center}
\end{figure}

\end{document}